\documentclass[usenatbib]{mn2e}

\usepackage{graphicx}
\usepackage{hyperref}

\newcommand\un[1]{{\,\rm #1}}
\newcommand\rs[1]{_\mathrm{#1}}
\newcommand\E[1]{\times10^{#1}}
\newcommand\apj{ApJ }

\newcommand\aap{A\&A}
\newcommand\g{$\gamma$}

\title[IC image of SN~1006]{Predicted gamma-ray image of SN~1006 due to inverse Compton emission}
\author[Petruk O., Bocchino F., Miceli M. et al.]{
O.~Petruk$^{1,2,4}$, F.~Bocchino$^{3,4}$, M.~Miceli$^{5,3,4}$, 
G.~Dubner$^{6}$, G.~Castelletti$^{6}$, \\ \\
{\LARGE \rm S.~Orlando$^{3,4}$, D.~Iakubovskyi$^{7}$, I.~Telezhinsky$^{8}$}\\
$^{1}$Institute for Applied Problems in Mechanics and Mathematics, Naukova St.\ 3-b,
  79060 Lviv, Ukraine\\
$^{2}$Astronomical Observatory, National University, Kyryla and Methodia St.\ 8, 79008 Lviv, Ukraine \\
$^{3}$INAF — Osservatorio Astronomico di Palermo, Piazza del Parlamento 1, 90134 Palermo, Italy\\
$^{4}$Consorzio COMETA, Via S. Sofia 64, 95123 Catania, Italy\\
$^{5}$Dipartimento di Scienze Fisiche ed Astronomiche, Sezione di 
Astronomia, Universit{\`a} di Palermo, \\ Piazza del Parlamento 1, 90134 Palermo, Italy\\
$^{6}$Instituto de Astronom\'{\i}a y F\'{\i}sica del Espacio (IAFE), CC 67, Suc. 28, 1428 Buenos Aires, Argentina\\
$^{7}$Bogolyubov Institute for Theoretical Physics, Metrologichna St. 14-b 03680 Kiev, Ukraine\\
$^{8}$Astronomical Observatory, Kiev National Taras Shevchenko University, Observatorna St. 3, 04053 Kiev, Ukraine\\
}
\begin{document}

\date{Accepted .... Received ...; in original form ...}

\pagerange{\pageref{firstpage}--\pageref{lastpage}} \pubyear{2002}

\maketitle

\label{firstpage}

\begin{abstract}
We propose a method to synthesize the inverse Compton (IC) gamma-ray image of a supernova remnant starting from the radio (or hard X-ray) map and using results of the spatially resolved X-ray spectral analysis. The method is successfully applied to SN~1006. We found that synthesized IC gamma-ray images of SN~1006 show morphology in nice agreement with that reported by the H.E.S.S. collaboration. The good correlation found between the observed very-high energy gamma-ray and X-ray/radio appearance can be considered as an evidence that the gamma-ray emission of SN~1006 observed by H.E.S.S. is leptonic in origin, though the hadronic origin may not be excluded. 
\end{abstract}


\section{Introduction}

H.E.S.S. and MAGIC projects open the era of the very-high energy (VHE) 
$\gamma$-ray astronomy in the sense that they  can provide for the first time
detailed  images of various astrophysical objects. 
VHE $\gamma$-ray emission from supernova remnants (SNRs) is one of the key 
components in investigation of the processes around strong nonrelativistic shocks, 
namely, dynamics and structure of the shock itself, magnetic field behavior, 
 and  microphysics of charged particles including their injection and 
acceleration. 
Observations in this energy domain  demonstrate  that charged particles 
are really accelerated in SNRs up to the highest energies 
observed in galactic cosmic rays. 

SNRs with known TeV $\gamma$-ray images include 
RX J1713.7-3946 \citep{RX1713aha2006,RX1713aha2007}, 
Vela Jr. \citep{VelaJraha2005,VelaJraha2007}, 
SN~1006 \citep{HESSproc2009}, 
RCW86 \citep{RCW862008}, 
IC443 \citep{IC443albert2007}, 
W28 \citep{W28aha2008}, 
CTB 37B \citep{CTB37B2008}, 
G0.9+0.1 \citep{G0901aha2005}, 
MSH 15-52 \citep{MHS1552aha2005}. 
Pulsar-wind nebulae could be the origin of $\gamma$-rays in G0.9+0.1 and  
MSH~15-52 while the particles accelerated at the forward shock of SNRs are 
likely to be  responsible  for emission in the other shell-type SNRs. 

The broad-band analysis of the spectra from SNRs  is useful to set 
constraints  on model parameters but it  still leave open the
nature of  VHE \g-ray flux either as leptonic or as hadronic in 
origin \citep[e.g. RX J1713.7-3946:][]{RX1713aha2006,RX1713Ber-Volk-06}. The 
analysis of the spatial distribution of \g-ray emission is an additional 
important channel of the experimental information. 

Hadronic \g-rays arise at the location of the target protons. 
Rather large density of target protons  -- as e.g. in 
molecular clouds -- 
is the condition for the effective hadronic emission in SNRs with 
high TeV \g-ray fluxes. The 
morphology of this type of emission in such SNRs is expected to follow the structures of 
regions of enhanced density of target protons, not the structures in 
the SNRs where initial protons are accelerated. 
In SNRs which are bright in $\gamma$-rays, surface brightness distribution of proton-origin \g-ray 
emission may not therefore be expected to follow the radio and/or nonthermal X-ray images of SNRs, like 
it is  observed  in IC443, W28 or CTB~37B. 

In contrast, the VHE \g-ray and hard X-ray morphologies  are observed to be
well correlated in the cases  of RX~J1713.7-3946, 
Vela~Jr., SN~1006 and possibly  RCW86.  
Thus it could be that TeV \g-rays reflect the same structures where the radio 
and nonthermal X-ray emission arise. 
In this scenario, electrons with energies of tens TeV may be responsible both 
for the (synchrotron) 
X-rays with energies of few keV and for the inverse-Compton 
(IC) \g-rays with energies of few TeV which are observable by H.E.S.S. 

The H.E.S.S. image of SN~1006 reported recently \citep{HESSproc2009}
reveals a very good correlation between X- and \g-ray maps. Can the 
correlation between IC \g-ray and synchrotron X-ray images really be an 
argument for leptonic origin of VHE $\gamma$-ray emission? 
In the present paper, we make use of the spatially resolved analysis of the radio 
and X-ray data of SN~1006  to generate  images 
 with the possible appearance that this SNR would acquire 
if the whole TeV \g-emission  were  due to leptonic IC process. 

 Since the purpose of the present analysis is to be as much model 
independent as possible, our work is mostly based on experimental results, 
without involving models of SNR dynamics, electron kinetics and evolution, 
etc., contrary to what has been carried out in previous approaches to the 
problem \citep{Reyn-98,reyn-fulbr-90,Reyn-04,Orletal07,thetak}. 
Radio  and nonthermal X-ray emission contain information about the 
accelerated electrons, their distribution inside SNR, maximum energies, etc. 
The method we propose extracts most of the important properties, which are 
needed to synthesize an IC \g-ray image, from the radio and X-ray data. The 
major exception is the magnetic field (MF).  
In the absence of observational information about it, we consider three cases 
of possible MF configurations. 


\section{Methodology}

Let us assume that the  energy  spectrum of electrons  holds the 
following relation 
\begin{equation}
 N(E)=KE^{-s}\exp(-E/E\rs{max}).
 \label{spectrelectons}
\end{equation}
where $N(E)$ is the number of electrons per unit volume with arbitrary directions of motion, $E$ the electron
energy, $K$ the normalization of the electron distribution, $s$ the power law index and $E\rs{max}$ the
maximum energy of electrons accelerated by the shock. 
This equation neglects small concave-up curvature of the spectrum predicted by efficient shock acceleration but allows us to be in the framework of the methodology of the X-ray spectral analysis \citep[][{\slshape srcut} model was used]{SN1006Marco}. 
The concavity results in a small bump around $E\rs{max}$ which leads mainly to some increase of the IC flux, which we are not interested in. It is not expected to affect the pattern of the gamma-ray brightness obtained with our method. 
Simpler spectrum $N(E)=KE^{-s}$ is valid for the radio emission. 
The emissivity due to synchrotron or IC emission is
\begin{equation}
 q(\varepsilon)=\int dE N(E) p(E,\varepsilon,[B])
 \label{ICpred:eq5} 
\end{equation}
where $p$ is the radiation power of  a  single electron with energy 
$E$, $\varepsilon$ is the photon  energy. 
The  strength of magnetic field $B$ is involved only in the synchrotron 
emission process. 	

The simplest way to reach our goal is to use the delta-function  approximation 
of the single-electron emissivities applied to spectrum 
 Eq.~(\ref{spectrelectons}). 
Namely, the special function $F$ appeared in the theory of synchrotron 
radiation is substituted with
\begin{equation} 
 {\cal F}\left(\frac{\nu}{\nu\rs{c}}\right)=
 \delta\left(\frac{\nu}{\nu\rs{c}}-0.29\right)
 \int\limits_{0}^{\infty}F(x)dx
 \label{F-delta}
\end{equation}
where $\nu$ is the frequency, $\nu\rs{c}(B,E)=c_1BE^2$ is the characteristic 
frequency, $c_1=6.26\E{18}\un{cgs}$. 
This results in synchrotron radio and X-ray emissivities
 with the following dependencies: 
\begin{equation} 
 q\rs{r}\propto\nu\rs{r}^{-(s-1)/2}KB^{(s+1)/2}
 \label{ICpred:eq1}
\end{equation}
\begin{equation} 
 q\rs{x}\propto\nu\rs{x}^{-(s-1)/2}KB^{(s+1)/2}
 \exp\left[-\left(\frac{\nu\rs{x}}{\nu\rs{break}}\right)^{1/2}\right]
 \label{ICpred:eq2} 
\end{equation}
where $\nu\rs{break}$ is 
\begin{equation} 
 \nu\rs{break}=\nu\rs{c}(B,E\rs{max}).
 \label{nurolloffdef}
\end{equation}
With  Eq.~(\ref{ICpred:eq1}) and Eq.~(\ref{ICpred:eq2}),  
we approximate the relation between the radio and X-ray synchrotron emissivities:
\begin{equation}
 q\rs{x}=q\rs{r}\left(\frac{\nu\rs{x}}{\nu\rs{r}}\right)^{-(s-1)/2}
 \exp\left[-\left(\frac{\nu\rs{x}}{\nu\rs{break}}\right)^{1/2}\right].
 \label{ICpred:eq3}
\end{equation}
However, exponentially cut off electron distribution 
 Eq.~(\ref{spectrelectons})  convolved with the $\delta$-function 
approximation for the single-particle emissivity, Eq.~(\ref{ICpred:eq2}), 
underestimates the synchrotron flux from the same electron distribution convolved 
with the full single-particle emissivity, Eq.~(\ref{ICpred:eq5}), at 
frequencies $\nu>30\nu\rs{break}$ 
(see Fig.~3 in \citet{Reyn-98}, $\nu\rs{break}$ is marked as $\nu\rs{m}$ there). 

In this paper, we use $\nu\rs{x}=2.4\un{keV}=5.8\E{17}\un{Hz}$. The range of 
$\nu\rs{break}$ in SN~1006 is found to be $(0.06\div 1)\E{17}\un{Hz}$ 
\citep{SN1006Marco}. Thus, we are working with 
$\nu\rs{x}\approx(6\div100)\nu\rs{break}$. Using 
Eq.~(\ref{ICpred:eq3}) , we may therefore underestimate the real X-ray flux 
in $\sim 10$ times in regions where $\nu\rs{break}$ is small. Thus, as it is 
pointed out by 
\citet{Reynolds-Keoh-99}, the approximate Eq.~(\ref{ICpred:eq3})  
is not robust at highest frequencies. 

We suggest therefore an empirical approximation of the numerically integrated 
synchrotron emissivity  Eq.~(\ref{ICpred:eq5}), i.e. emissivity of the 
exponentially cut off electron distribution 
convolved with the full single-particle emissivity:
\begin{equation}
 q\rs{x}=q\rs{r}\left(\frac{\nu\rs{x}}{\nu\rs{r}}\right)^{-(s-1)/2}
 \exp\left[-\beta\rs{x}\left(\frac{\nu\rs{x}}{\nu\rs{break}}\right)^{0.364}\right]
 \label{ICpred:eq4}
\end{equation}
where $\beta\rs{x}=1.46+0.15(2-s)$. This approximation is quite accurate. Its 
errors are less than $18\%$ for $s=1.8\div2.5$ and 
$\nu\rs{x}\leq 10^3\nu\rs{break}$. 

We assume that VHE \g-ray emission from SN~1006 is due to IC process in the 
black-body photon field of the cosmic microwave background. The convolution of 
the electron distribution  Eq.~(\ref{spectrelectons})  
with the $\delta$-function 
approximation for the single-particle IC emissivity \citep{Pet08IC} 
is also inaccurate  to describe the \g-ray radiation of electrons with 
energies around $E\rs{max}$. Therefore, like in the case of the synchrotron 
radiation, we use an approximate formula for the IC emissivity, too. 

\begin{figure}
 \centering
 \includegraphics[height=6.7truecm]{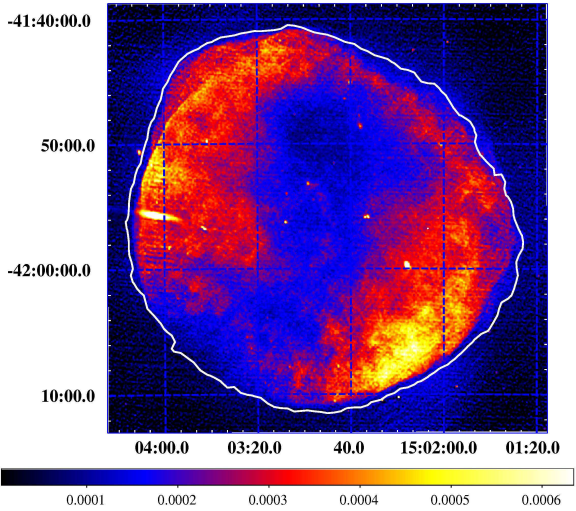}
 \caption{
 Radio image of SN~1006 at $\lambda\sim 20$ cm. The color scale is in units Jy beam$^{-1}$.
 The white contour denotes the boundary of SN 1006 (see text).
        }
 \label{ICpred:radio}
\end{figure}

Let us consider IC emission at $1\un{TeV}$. We developed an approximation for the 
numerically integrated IC emissivity (\ref{ICpred:eq5})  of 
electrons with the energy spectrum (\ref{spectrelectons})  at \g-ray 
energy $1\un{TeV}$: 
\begin{equation}
 q\rs{ic@1}\propto 
 K\exp\left(-{\beta\rs{ic}}{E\rs{max}^{-0.75}}\right)
 \label{ICpred:eq6}
\end{equation}
where $\beta\rs{ic}=15$ for $2\leq s\leq2.5$ and $\beta\rs{ic}=15+2(2-s)$ for 
$1.8\leq s<2$. The error of this approximation is less than $25\%$ for 
$E\rs{max}\geq0.3\un{TeV}$. This approximation accounts  for  the 
Klein-Nishina decline where necessary. 

The maximum energy is related to $\nu\rs{break}$ by 
 Eq.~(\ref{nurolloffdef}) :
\begin{equation}
 E\rs{max}=C_1 \nu\rs{break}^{1/2}B^{-1/2}
 \label{ICpredeq5}
\end{equation}
where $C_1=c_1^{-1/2}$. 
Substitution of  Eq.~(\ref{ICpred:eq6})  with this $E\rs{max}$ and $K$ 
from  Eq.~(\ref{ICpred:eq1})  results in 
\begin{equation}
 q\rs{ic@1}\propto q\rs{r} B^{-(s+1)/2}
 \exp\left[-\beta\rs{ic}\left(\frac{B^{1/2}}{C_1\nu\rs{break}^{1/2}}\right)^{0.75}\right].
 \label{ICpred:eq7}
\end{equation}
This expression relates the radio emissivity and the IC \g-ray emissivity at 1 
TeV  with only  one unknown, $B$. 
It may be used in the same cases where the {\slshape srcut} model is 
applicable, that is,  if the spectrum of electrons may be approximated 
by Eq.~(\ref{spectrelectons}) with $s$  assumed  constant from the radio to 
X-ray emitting electrons. 


\begin{figure}
 \centering
 \includegraphics[height=6.7truecm]{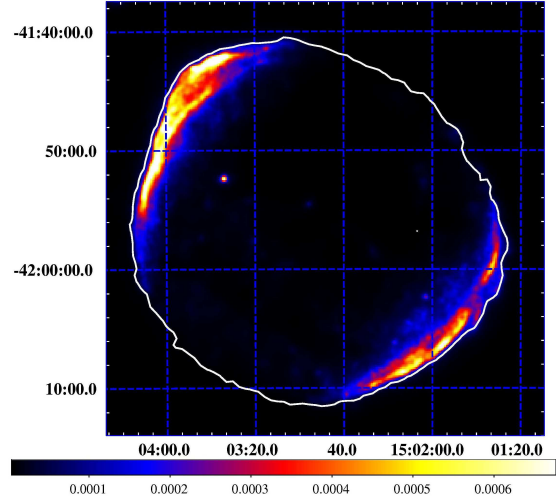}
 \caption{
 X-ray image of SN~1006 in $2-4.5\un{keV}$. 
 The pixel size is $8''$. 
 The color scale is in units MOS1 counts s$^{-1}$ pixel$^{-1}$. 
        }
 \label{ICpred:Xray}
\end{figure}

The idea of our method is represented by Eq.~(\ref{ICpred:eq7}). Namely, 
this expression may be used in a small region (``pixel'') of  a  
SNR  projection in  order to relate the surface brightness in radio band and 
IC $\gamma$-rays. 
Having this relation applied to all ``pixels'', we may predict the main 
features of the \g-ray morphology of SNR originated in an  IC process.
This procedure `converts' the radio image to  an  IC one. 

Another possibility is to start from the hard X-ray map and `translate' it into the 
\g-ray image in a similar fashion. 
Namely, substitution  Eq.~(\ref{ICpred:eq7})  with $q\rs{r}$ from 
 Eq.~(\ref{ICpred:eq4})  yields 
\begin{equation}
\begin{array}{l}\displaystyle
 q\rs{ic@1}\propto q\rs{x}B^{-(s+1)/2}
 \exp\left[-\beta\rs{ic}\left(\frac{B^{1/2}}{C_1\nu\rs{break}^{1/2}}\right)^{0.75}
 \right.\\ \\ \displaystyle \left. \qquad
 +\beta\rs{x}\left(\frac{\nu\rs{x}}{\nu\rs{break}}\right)^{0.364}\right].
\end{array} 
 \label{ICpredeq8}
\end{equation}
However, an X-ray image  can  be used  only  if it is dominated 
everywhere by the nonthermal emission. 

\begin{figure*}
 \centering
 \includegraphics[width=6.6truecm,angle=90]{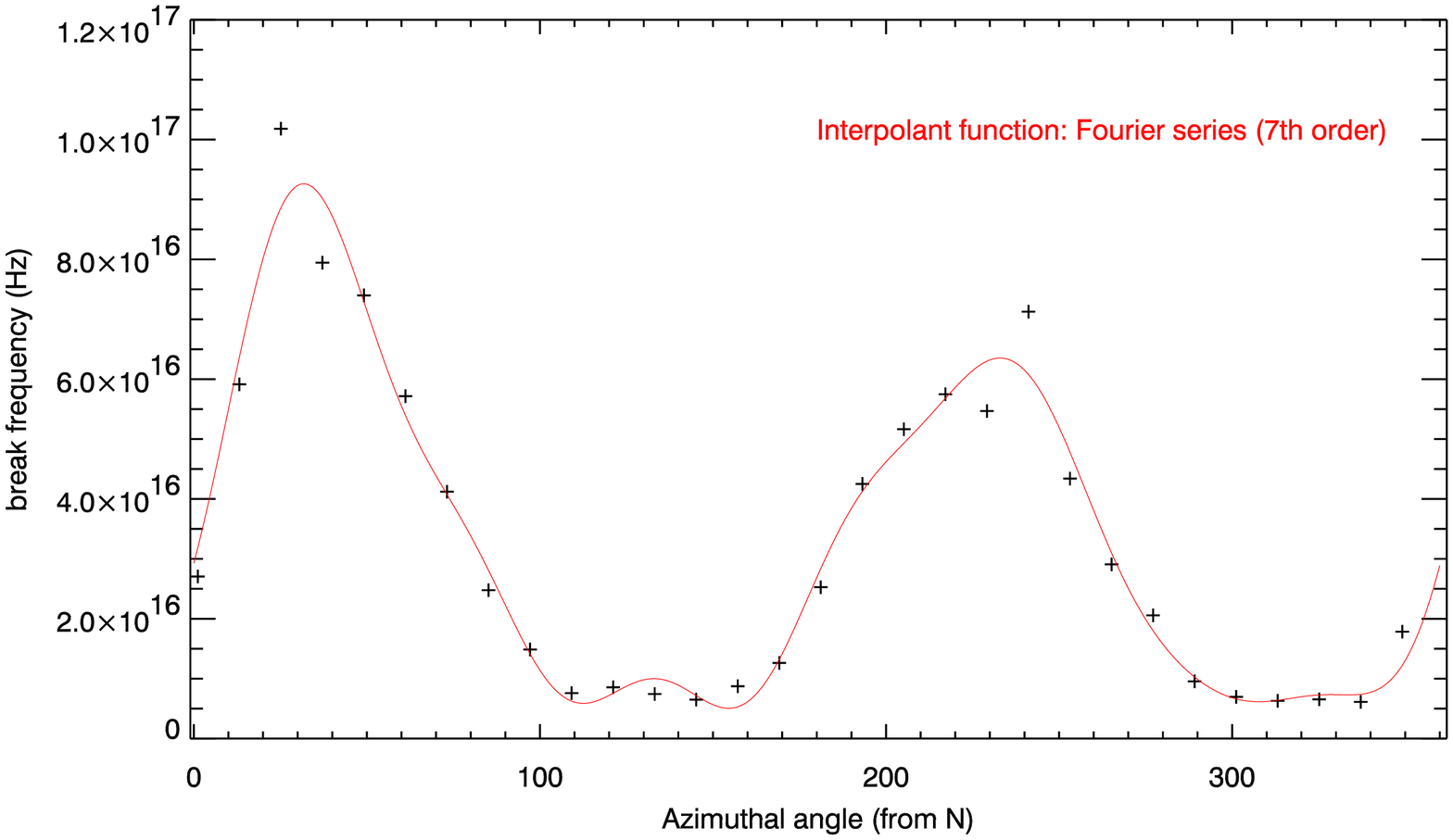}
 \includegraphics[height=6.2truecm]{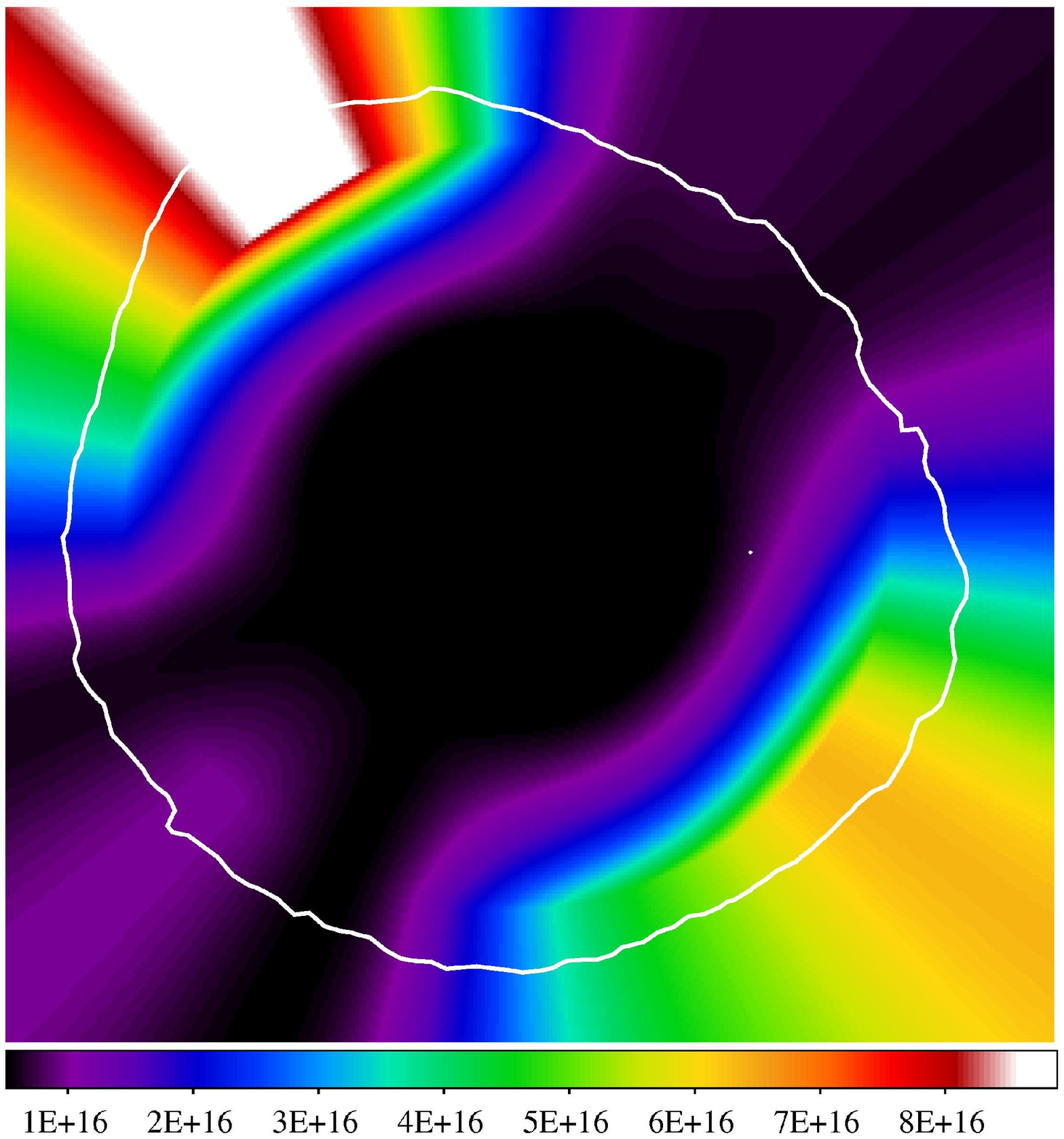}
 \caption{Break frequency $\nu\rs{break}$: azimuthal profile (left) and image 
(right, the color scale is in units of Hz). 
        }
 \label{ICpred:nurolloff}
\end{figure*}

Eqs.~(\ref{ICpred:eq7}) and (\ref{ICpredeq8}) relate emissivities of the uniform plasma. 
We use these equations to deal with surface brightnesses that are superpositions of the local emissivities 
along the line of sight. 
Strictly speaking, this may be done only for the thin rim around SNR edge where plasma is 
approximately uniform along the line of sight. 
However, this approach may also be extended to deeper regions of 
SNR projection (see Appendix).
Since X-ray limbs are (and \g-ray ones are expected to be) quite thin and close to the edge, our method is able 
to correctly determine the location of the bright limbs in the IC \g-ray image of SN~1006. 
In the interior of SNR projection, we consider $B$ and $\nu\rs{break}$ as `effective' values 
for a given `pixel'. 

It is interesting to note that Eq.~(\ref{ICpred:eq7}) [or Eq.~(\ref{ICpredeq8})] 
may be solved for the value of $B$. 
In this way, the method proposed here may be used for deriving the 
effective (line-of-sight averaged) MF pattern in SNR from its radio 
[or synchrotron X-ray] and IC \g-ray maps. The distribution of 
$\nu\rs{break}$ may be obtained 
from the radio ($q\rs{r}$) and synchrotron X-ray ($q\rs{x}$) images by 
solving Eq.~(\ref{ICpred:eq4}), without the need of the spatially resolved X-ray analysis. 

\begin{figure}
 \centering
 \includegraphics[width=6truecm]{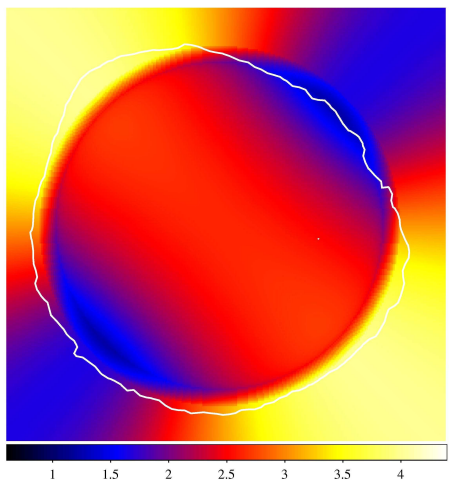}
 \caption{Map of effective MF used in calculations for the model MF1. 
 $\zeta=0.9$. 
 This map is rotated on $90^\mathrm{o}$ for the other scenario, MF2. 
 The color scale is in units of $B\rs{o}$. 
        }
 \label{ICpredict:figMF}
\end{figure}

\section{Experimental data and models of magnetic field}

In order to make use of  Eq.~(\ref{ICpred:eq7})  or 
 Eq.~(\ref{ICpredeq8}), one needs: 
i) an initial image, i.e. the map of distribution of the synchrotron radio 
(or X-ray) surface brightness, ii) the distribution of $\nu\rs{break}$ 
obtained from the spatially resolved spectral analysis of X-ray data and iii) 
the distribution of the effective magnetic field over the initial image. 

We use the  high resolution  radio image of SN~1006 at $\lambda\sim 20$ 
cm (Fig.~\ref{ICpred:radio}), produced on the basis of archival Very Large 
Array data combined with Parkes single-dish data 
 presented and described in \citet{SN1006mf}. 

As for the X-ray image, we use the 2.0-4.5 keV XMM-Newton EPIC
  mosaic obtained by \citet{SN1006Marco}, shown in Fig.~\ref{ICpred:Xray}. 
  The mosaic has been produced by combining the 7 public archive
  XMM-Newton observations of SN1006, and using the PN, MOS1 and MOS2
  cameras. The total effective EPIC exposure times ranges between 70 and 140
  MOS1-equivalent ks. 

A contour plotted on  the  images  delineates  the boundary of 
SN~1006. It corresponds to the level of $10\%$ of the maximum 
brightness in the soft ($0.5-0.8$ keV) X-ray map of SNR 
 \citep[see Fig.~1 in][]{SN1006Marco}. 

The azimuthal profile (Fig.~\ref{ICpred:nurolloff}a) and image 
(Fig.~\ref{ICpred:nurolloff}b) of the break frequency was derived on the basis 
of spatially resolved X-ray spectral fitting results. The reader is referred to 
\citet{SN1006Marco} for the details on the procedure for X-ray data analysis and 
creation of this image. 
The same spectral fits also show that  $\alpha$, the radio spectral 
index,  is between 0.47 and 0.53 everywhere around the shock. 
Therefore, we take $s=2\alpha+1=2$ to be constant in SN~1006. 

We consider three models for magnetic field inside SNR. 
SN~1006 is rather symmetrical. In the procedure of MF map simulation, SNR is 
assumed to be spherical, with the radius equal to the average radius of SN~1006. 
For technical reasons, MF is fixed to the postshock value also outside the 
boundary of the spherical SNR (Fig.~\ref{ICpredict:figMF}). This allows us to 
deal correctly also with regions of SN~1006 which are larger than the average 
radius. 

Classical MHD description corresponds to the unmodified shock theory. 
It takes into account the post-shock evolution of MF and the compression 
factor which increases with the shock obliquity.
In this case, two possible orientations of ISMF  are  
considered. Namely, NW-SE (Fig.~\ref{ICpredict:figMF}) in model MF1 
 (equatorial, or barrel-like, model)  and NE-SW in MF2 
 (polar caps model). 
ISMF is assumed to be constant around SN~1006 with the strength 
$B\rs{o}=10\un{\mu G}$. 
This value is choosen to give, in models MF1 and MF2, the postshock magnetic 
field $20\div40\ \mu G$, a value which follows from estimations for downstream MF strength 
\citep[e.g.][]{volk-ber-ksen2008} and reported \g-ray flux \citep{HESSproc2009}.
Such ISMF looks to be unrealistic at the position of SN 1006 far above the Galactic plane. 
A possibility to provide tens of $\mu G$ upstream of the shock would be the magnetic field amplificaion 
as an effect of efficient cosmic ray acceleration, which is out of the scope of this study. 
It should be noted however that the absolute value of the upstream field plays no role 
for the purpose of our paper. 
The aspect angle between ISMF and the line of sight 
is taken to be $70^\mathrm{o}$ for MF1 \citep{Reyn96,SN1006mf} and, 
for simplicity, also for MF2 (we shall see below that the actual value of the aspect 
angle is not crucial for the purpose of the present paper, because even 
different models of MF lead to quite similar \g-ray pattern).

\begin{figure}
 \centering
 \includegraphics[width=7truecm]{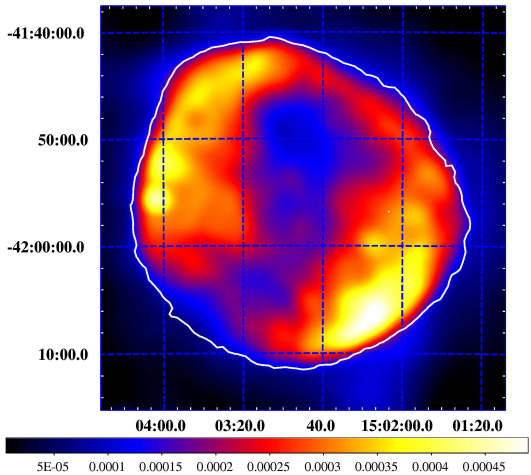}
 \caption{Radio image of SN~1006 at $\lambda\sim 20$ cm, from 
 Fig.~\ref{ICpred:radio}, smoothed with Gaussian 
 with $2'$ sigma. 
        }
 \label{ICpred:Srsmothed-to-HESS}
\end{figure}

The procedure of generation of the average MF maps for MF1 and MF2 models is as follows.
First we calculated numerically MHD model of Sedov SNR \citep{Sedov-59,Reyn-98,Korob1991}. 
This gives us three-dimensional distribution of MF inside SNR. 
An effective MF in a given `pixel' of SNR projection is taken as a straight average 
for this 3-D MF distribution along the line of sight, accounting for the azimuthal 
orientation and an aspect angle of the ambient MF in respect to the observer 
as well as the fact that most of emission arise right after the shock.  
Really, in order to generate map of effective MF one should calculate 
the emissivity-weighted average. 
Such an approach requires however the knowledge of 3-D distribution of emitting 
electrons within SNR which is unknown until one makes the full modelling from basic 
theoretical principles. 
In contrast, the scope of the present paper is to `extract' 
the structure of radiating material from the observational data. 
Therefore, our procedure consists in approximate calculation of emissivity-weighted MF, 
without considering the 3-D distribution of relativistic electrons. In fact, 
most emission comes from  a  rather thin shell with thickness 
$\sim 10\%$ of SNR radius. 
Therefore, in calculations of the average magnetic field, we consider  
only this part of SNR interior, namely the integration along the line of sight is within regions 
from $\zeta R$ to $R$, with $\zeta<1$ (Fig.~\ref{ICpredict:figMF}). 
A choice of $\zeta$ is rather arbitrary. It is apparent from calculations 
that contrasts between the outer regions and the interior of the IC image depend on 
this choice.
The preference to the value of $\zeta$ does not alter, however, the main 
features of the predicted IC morphology of SNR. Note that the azimuthal 
variation of the brightness is not affected at all by $\zeta$ for radii of SNR 
projection $\geq \zeta R$. 
In other words, the position of the limbs on the synthesized IC \g-ray image may not 
be altered by the particular choice of $\zeta$. 
Nevertheless, in order to determine the most appropriate value of $\zeta$, we made the full 
MHD simulations of Sedov SNR with model of evolution of the relativistic electrons in the SNR interior 
from \citet{Reyn-98}. Then the map of the emissivity-weighted average MF was produced from these simulations 
and compared with our maps of effective MF derived for different $\zeta$. In this way, we found that the value 
$\zeta=0.9$ provides the good correspondence between MF maps in these two approaches.

The third model of MF is relevant to the nonlinear acceleration theory with the 
time-dependent MF amplification and the high level of turbulence \citep[Bohm limit;][]{volk-ber-ksen2004}. 
The quasi-parallel theory assumes in this 
case that the turbulence is produced ahead of the shock, not downstream. The compression 
of the (already turbulent) magnetic field then does not depend on the original obliquity 
\citep{Volk-Ber-injobliq-2003,Ber-Ksenof-SN1006-2002}. \citet{Rakowski-2008} argue 
that shocks of different initial obliquity 
subject to magnetic field amplification become perpendicular immediately 
upstream.
ISMF is therefore assumed to increase on the shock by a large factor 
(due to compression and amplification), the same for any obliquity.
In addition, in model MF3, we assume MF to be approximately uniform everywhere inside 
SNR \citep{ber-volk2004-SD}, with the strength $150\un{\mu G}$ \citep{BVK2005}. 

Theoretical work on magnetic-field
amplification starting with \citet{bell-lucek2001} focuses on shocks which 
are originally parallel far upstream, with 
some implications that the process is less effective for (initially) perpendicular 
shocks. In this scenario, obliquity dependence of the post-shock MF would be  
opposite to the classical one. Really, MF amplification is expected to follow 
acceleration efficiency which decrease with the obliquity. If so, limbs in SN~1006 
should correspond to the largest post-shock MF. Such MF morphology is qualitatively 
represented by MF1 model (Fig.~\ref{ICpredict:figMF}). 

\begin{figure}
 \centering
 \includegraphics[width=7truecm]{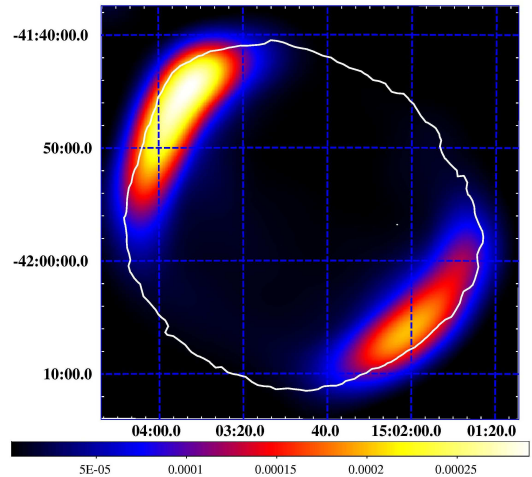}
 \caption{X-ray image of SN~1006 in 2-4.5$\un{keV}$, from 
 Fig.~\ref{ICpred:Xray}, smoothed with Gaussian 
 with $2'$ sigma. 
        }
 \label{ICpred:Sxsmothed-to-HESS}
\end{figure}

Some other notes to our methodology are in order. 
All initial maps were  homogenized to the  
same size, orientation, resolution and pixel size. Eq.~(\ref{ICpred:eq7}) is 
applied to each pixel of the initial images. 
The maximum brightness on images is fixed at the maximum value in the histogram distribution which 
has at least 10 pixels. The minimum is fixed at the level 1/100 of the maximum 
value. This is true for all the images, including the synthesized images, 
the observed X-ray and radio ones, and excluding the MF and $\nu\rs{break}$ maps. 

Some images are smoothed to fit the resolution of H.E.S.S. 
The resolution used is $\mathrm{FWHM}=4.75'$ \citep{HESSresolution}, so the Gaussian sigma 
is $2'$. 
The role of smoothing is visible on Figs.~\ref{ICpred:Srsmothed-to-HESS}, \ref{ICpred:Sxsmothed-to-HESS}, 
to be compared with radio and X-ray images on Figs.~\ref{ICpred:radio}, \ref{ICpred:Xray}.

\begin{figure}
 \centering
 \includegraphics[width=6truecm]{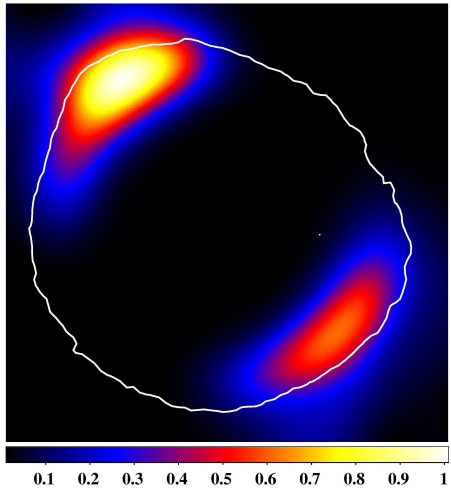}
 \caption{X-ray image of SN~1006 at $2.4\un{keV}$ generated from 
 the radio one (Fig.~\ref{ICpred:radio}) 
 with the use of Eq.~(\ref{ICpred:eq4}) and 
 smoothed to $2'$ Gaussian sigma. 
 The color scale is normalized to the maximum brightness. 
        }
 \label{ICpred:SxfromSr}
\end{figure}

\section{Results and Discussions}
\label{icp:discussion}

To  check our approach,  we made use of 
Eq.~(\ref{ICpred:eq4})  to generate the X-ray image of SN~1006 starting 
from the radio one.  We stress that we do not need any assumption about MF 
configuration for this test, everything needed for radio to X-ray 
conversion may be taken from observations. Really, we need just the input  
radio image (Fig.~\ref{ICpred:radio}), and 
the distribution of the break frequency (Fig.~\ref{ICpred:nurolloff}).  The 
resulting  image is presented on Fig.~\ref{ICpred:SxfromSr}. It shows 
good correlation  with the hard X-ray observations 
(Fig.~\ref{ICpred:Sxsmothed-to-HESS}), confirming  that the proposed method works well. 
It restores the main properties of the observed hard X-ray image: essential 
decrease of the thickness of the two bright synchrotron limbs in X-rays 
comparing to the radio band (Fig.~\ref{ICpred:Srsmothed-to-HESS}), 
correct position of the limbs
and negligible emission from  the interior.  
Therefore, the method  is reliable to  be used for 
simulation of the \g-ray images of SNRs\footnote{%
The agreement between the observational and the synthesized X-ray maps is not perfect because 
the `effective' resolution of the synthesized X-ray image cannot be better than the `resolution' 
of the image of $\nu\rs{break}$. 
The effective resolution of $\nu\rs{break}$ image is defined by the size of 
30 rim regions used for spectral analysis \citep[see Fig.~1 in][]{SN1006Marco}
that in turn is determined by the photon statistics. 
Namely, 
the {\it radial} resolution in the image of $\nu\rs{break}$ is limited, since the radial profile of the cut-off frequency is assumed to be constant inside the rim regions. Note that the {\it azimuthal} `resolution' is better 
than radial (see profile of $\nu\rs{break}$ on Fig. 3).}. 

\begin{figure*}
 \centering
 \includegraphics[width=5.8truecm]{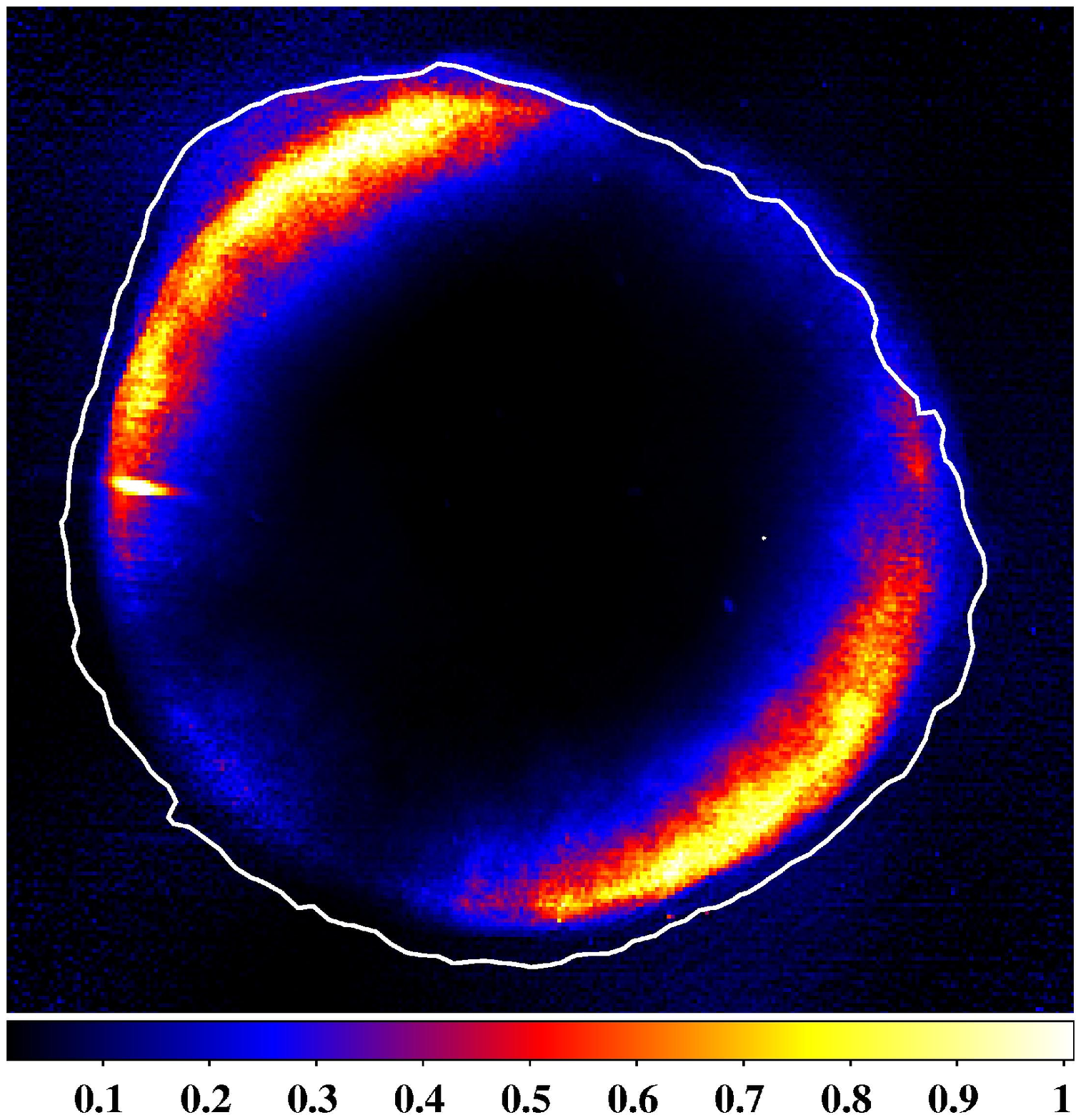}
 \includegraphics[width=5.8truecm]{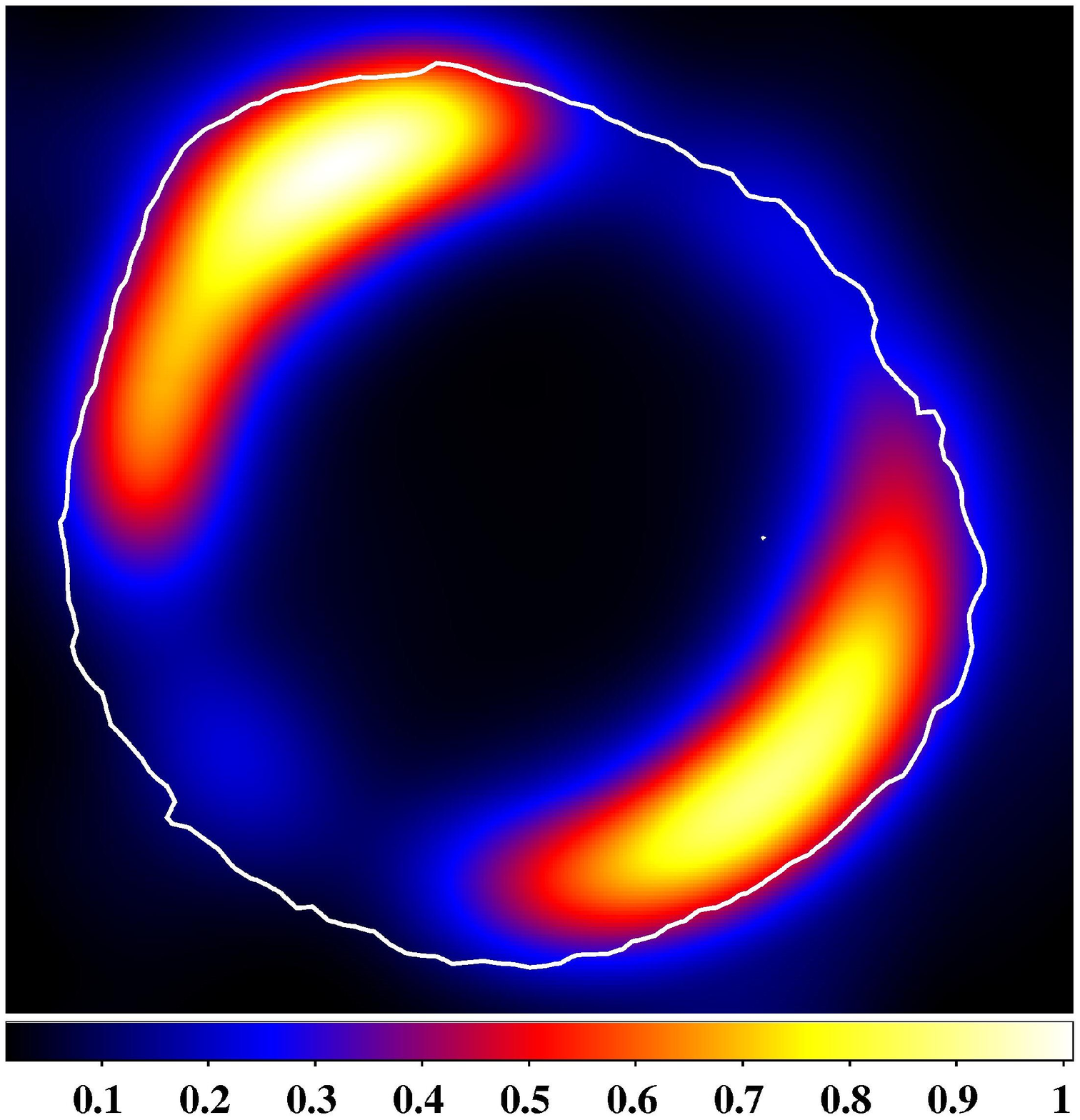}
 \includegraphics[width=5.8truecm]{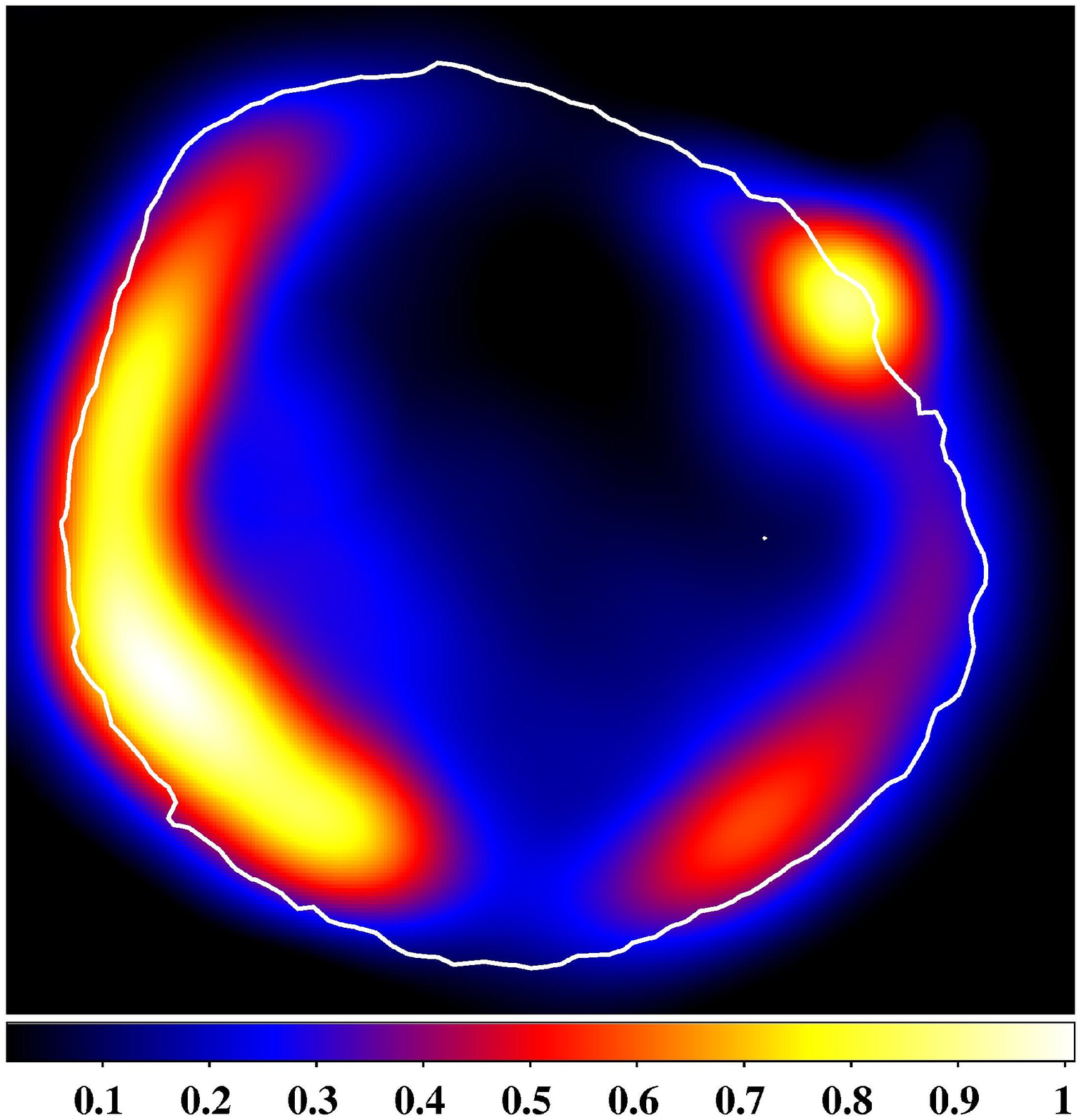}
 \caption{Predicted IC morphology of SN~1006 at photons with energy 
 $1\un{TeV}$, for the model MF1. 
 IC image generated from the radio map (left); 
 the same smoothed to $2'$ Gaussian sigma to fit the H.E.S.S. 
 resolution (centre); 
 IC image generated from X-ray map and smoothed to $2'$ Gaussian sigma (right).
 The color scales are normalized to the maximum brightness.
        }
 \label{ICpred:ICimages:MFa}
\end{figure*}

\begin{figure*}
 \centering
 \includegraphics[width=5.8truecm]{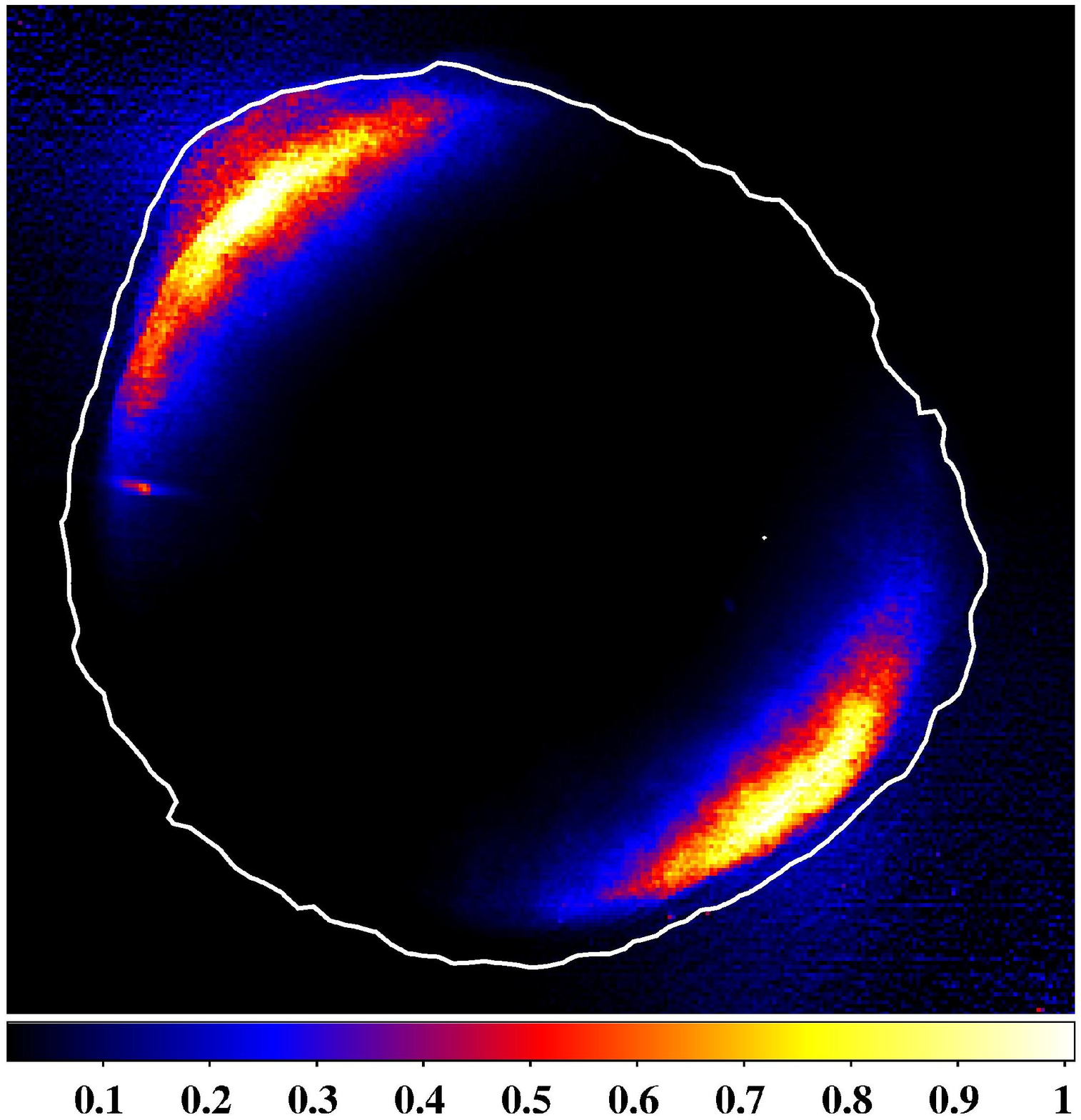}
 \includegraphics[width=5.8truecm]{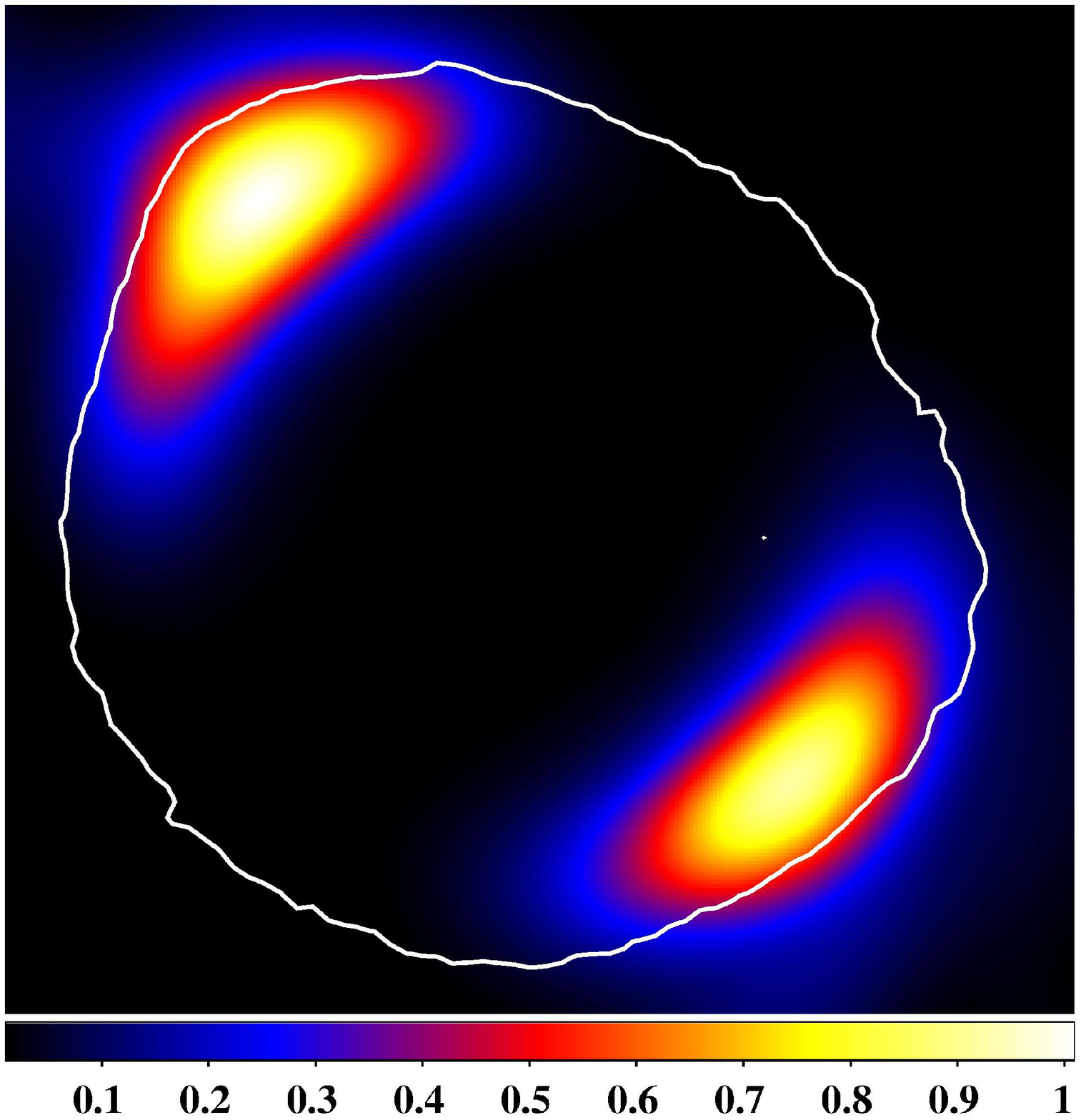}
 \includegraphics[width=5.8truecm]{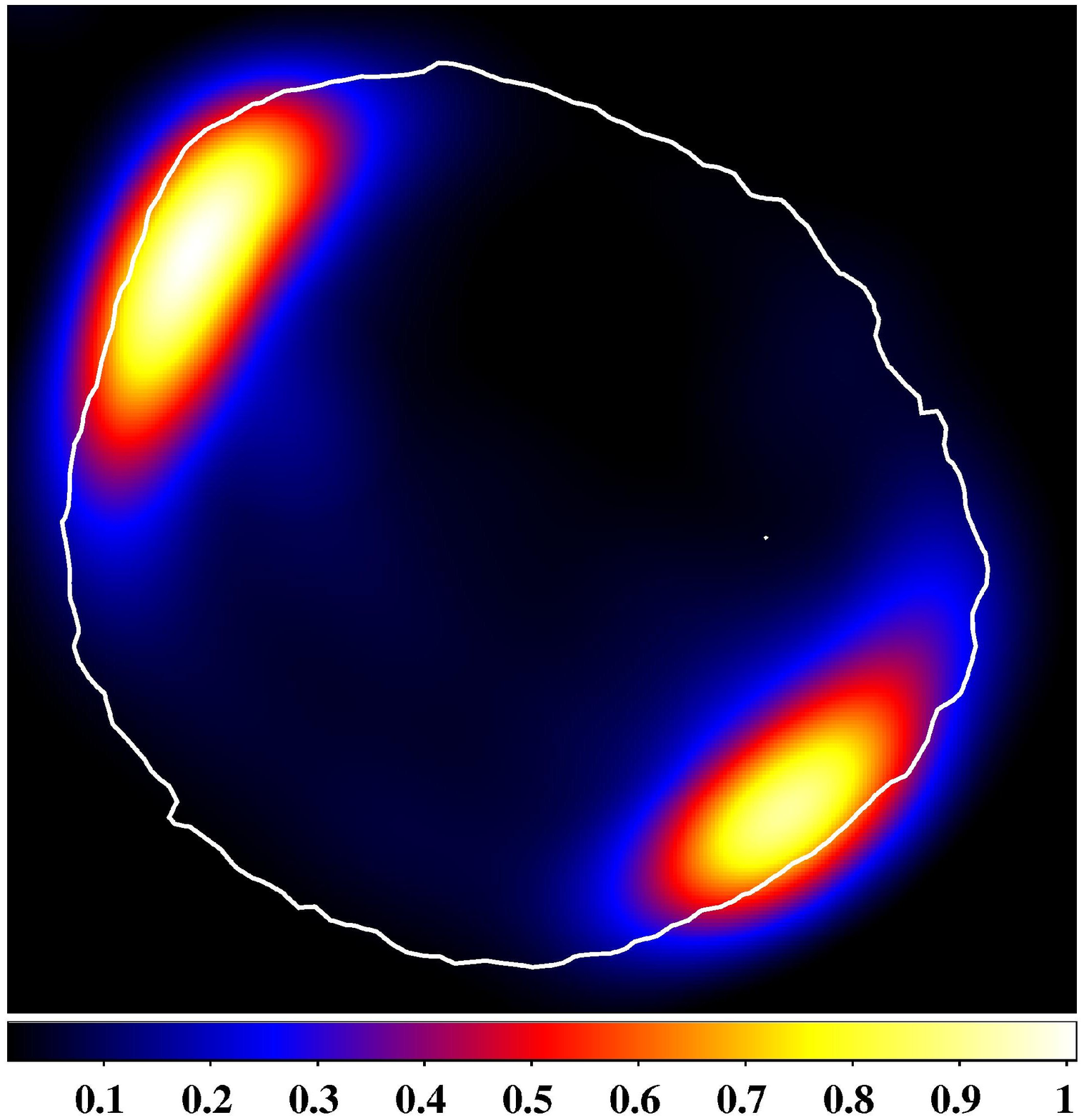}
 \caption{The same as on Fig.~\ref{ICpred:ICimages:MFa} for the model MF2.
         }
 \label{ICpred:ICimages:MFb}
\end{figure*}

\begin{figure*}
 \centering
 \includegraphics[width=5.8truecm]{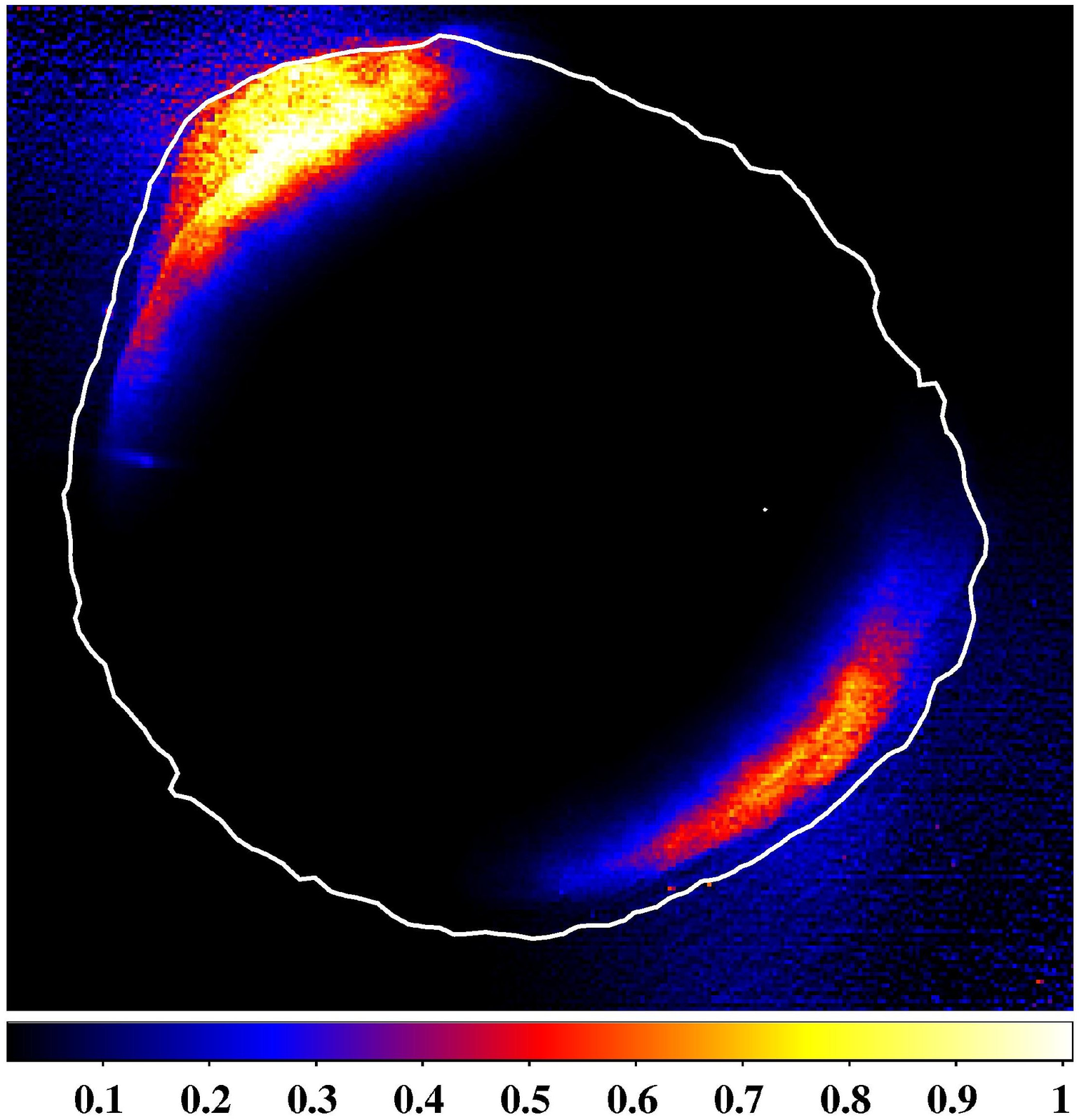}
 \includegraphics[width=5.55truecm]{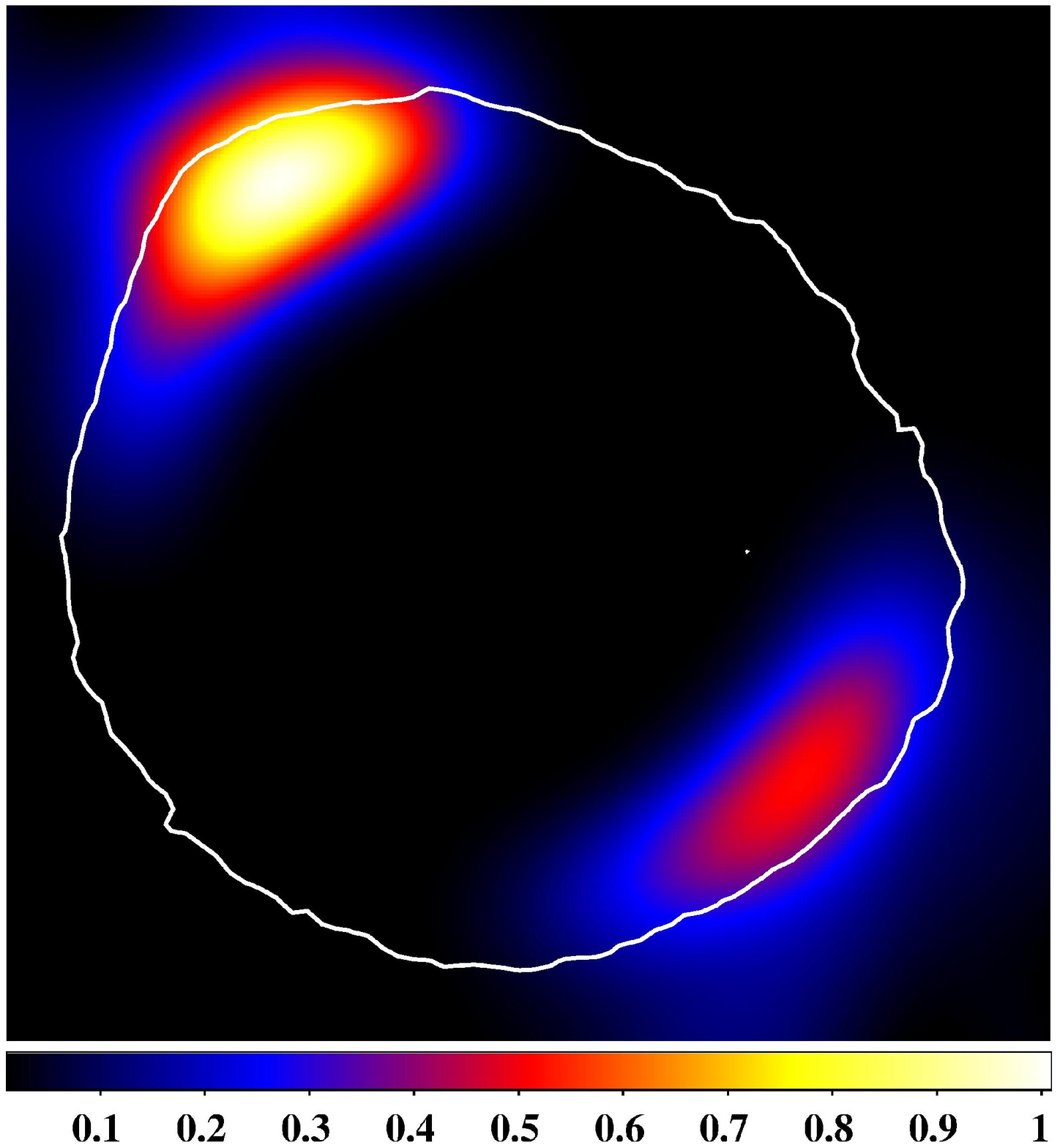}
 \includegraphics[width=5.8truecm]{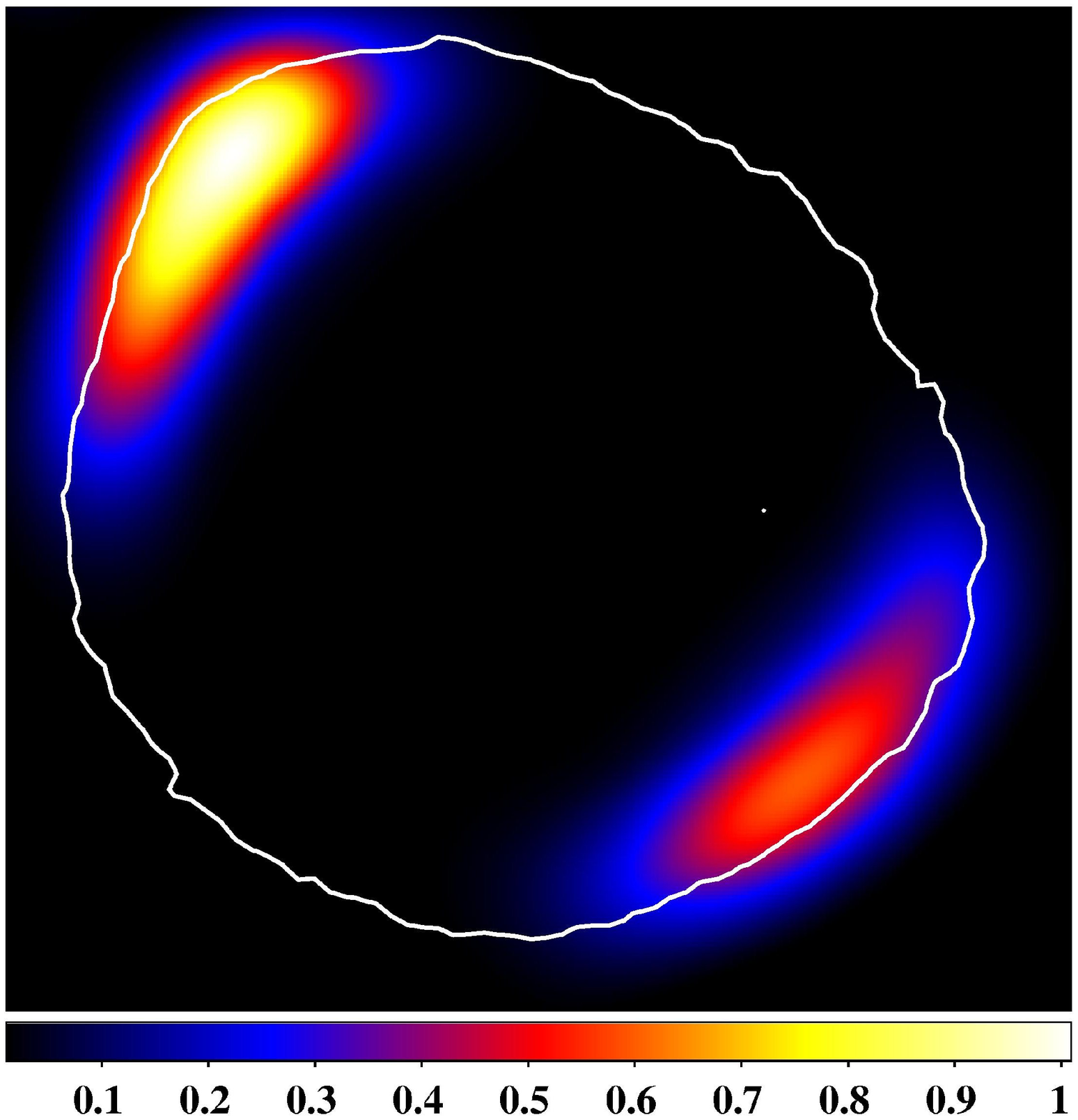}
 \caption{The same as on Fig.~\ref{ICpred:ICimages:MFa} for the model MF3.
         }
 \label{ICpred:ICimages:MFc}
\end{figure*}

Synthesized TeV \g-ray images of SN~1006 due to IC process are presented on 
Fig.~\ref{ICpred:ICimages:MFa} (model MF1, barrel-like SNR in classical MHD or polar caps in 
non-linear \citet{bell-lucek2001} approach), Fig.~\ref{ICpred:ICimages:MFb} 
(MF2, polar-caps in classical MHD) and Fig.~\ref{ICpred:ICimages:MFc} (MF3, uniform MF in the SNR interior for \citet{Ber-Ksenof-SN1006-2002} non-linear model).  
Images presented on the
left panels were  obtained from the radio map as initial 
model,  while those shown in the right panels have the hard X-ray map 
as the starting point.  Middle panels represent `radio-origin' \g-images 
of the left panels smoothed to the resolution of H.E.S.S. 

Let us  first consider the \g-ray morphologies obtained from the radio image. 
Two  arcs dominate in all three MF configurations. Their locations 
correspond to limbs in radio and X-ray images. This confirms that correlation 
between TeV \g-ray and X-ray/radio morphologies may be considered as direct 
evidence that the \g-ray emission of SN~1006 observed by H.E.S.S. is leptonic in 
origin. 

Geometry of MF essentially different from those we considered 
might result in different predicted \g-ray images of SNR. 
Nevertheless, our results demonstrate that if MF strength varies within 
factor $\sim 3$ around the shock (Fig.~\ref{ICpredict:figMF}), 
any configuration of MF results in double-limb IC \g-ray image of SN~1006. 


For areas with the same radio surface brightness, higher MF implies lower IC \g-ray 
brightness, Eq.~(\ref{ICpred:eq7}). We 
have therefore different azimuthal contrasts (i.e. the ratio 
of maximum to minimum brightness around the rim) 
in  \g-ray  images for different models of MF. 
In particular, in model MF1 the strength of the 
postshock MF is maximum in NE and SW regions (Fig.~\ref{ICpredict:figMF}) where 
the radio brightness has maxima. At the same time, MF is smaller in faint NW and SE regions. 
This leads to small azimuthal contrast of brightness 
between bright and faint regions in \g-ray image (Fig.~\ref{ICpred:ICimages:MFa}, left). 
In the opposite model MF2 the strength of 
the postshock MF is maximum in the faint NW and SE regions, 
that results in the largest brightness contrast 
(Fig.~\ref{ICpred:ICimages:MFb}, left). 
Two arcs are therefore more pronounced in the \g-ray image for models MF2 and MF3. 

The differences in \g-images for three models of MF are not so prominent 
 after smoothing  them to the resolution of H.E.S.S. In all three cases 
(Fig.~\ref{ICpred:ICimages:MFa}-\ref{ICpred:ICimages:MFc}, center panels), 
there are two bright limbs in the same  locations corresponding  to 
limbs on the smoothed X-ray map shown  in  
Fig.~\ref{ICpred:Sxsmothed-to-HESS}. The synthesized images  can be also 
directly compared to the H.E.S.S. map of  SN~1006 \citep{HESSproc2009}.
Good correlation between the synthesized and the observed images allows us 
to prefer leptonic origin of TeV \g-ray emission of this SNR. 
The uncertainties introduced in our method by the lack of knowledge of the real
MF inside the remnant do not alter this correlation, as it is clearly seen from 
our synthesized images. 

VHE \g-ray images obtained from the initial hard X-ray map 
(Fig.~\ref{ICpred:ICimages:MFa}-\ref{ICpred:ICimages:MFc}, right panels) are 
quite similar to those from the initial radio map, except for the configuration 
MF1. 
This fact reinforce the goodness of our method.
The reason of differences between the `radio-origin' and the `X-ray-origin' 
\g-ray images in the MF1 case (Fig.~\ref{ICpred:ICimages:MFa}right) 
is the contribution of the thermal X-ray 
emission to the hard X-ray image which was used, in the SE and NW regions of SN~1006. 
Namely, our  fitting show that the fraction of thermal emission in the overall 2-4.5 keV flux 
in the SE region is about 50\% \citep{SN1006Marco}. 
The prominent but localized NW bright spot is completely dominated by the thermal X-ray emission 
\citep{Vink_et_al2003}.
So, the \g-ray brightness in these 
regions is overestimated in our synthesized images. This effect is not prominent 
for MF2 and MF3 configurations because MF is large enough in SE region to 
visually decrease the brightness there. 
Thus, the X-ray map may be used as initial one in our method only 
if it is completely dominated everywhere by the nonthermal emission. 

\begin{figure}
 \centering
 \includegraphics[width=8.4truecm]{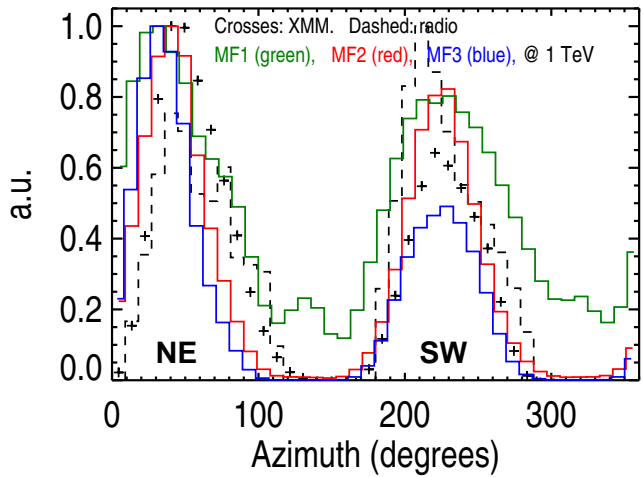}
 \caption{Azimuthal profiles of the observed radio (dashed line) 
 and X-ray (crosses) compared with 
 the synthesized \g-ray (solid lines) surface brightness 
 (as from the input radio image), 
 from images smoothed with $2'$ sigma Gaussian. 
 Background galaxy at the azimuth around $90^o$ is excluded.
        }
 \label{ICpred:profiles}
\end{figure}

In order to quantify differences in images due to different models of MF, 
the azimuthal profiles from our synthesized \g-ray images are compared on Fig.~\ref{ICpred:profiles} 
with the corresponding azimuthal profiles derived from 
the VLA+Parkes radio image (Fig.~\ref{ICpred:Srsmothed-to-HESS}) and 
the XMM-Newton X-ray image of SN~1006 (Fig.~\ref{ICpred:Sxsmothed-to-HESS}). 
The profiles have been derived from maps smoothed to the HESS resolution.  
A large (from $8'$ to $20'$ from center) annulus 
centered on the remnant is divided in 40 sectors. The values of brightness plotted 
is obtained by integration inside the sectors. 
Observational profiles are corrected for the background which is determined 
inside the central $0'-7'$ circle. 

Fig.~\ref{ICpred:profiles} shows that the radio azimuthal profile may not be used as a tracer 
for the azimuthal variation of the IC \g-ray brightness because it reflects only the changes in 
MF strength and density of relativistic electrons, while X-ray profile accounts also for the 
variation of the electron maximum energy which is important for \g-ray emission as well. 

Fig.~\ref{ICpred:profiles} reveals a very good match between positions and shape of the limbs in 
the hard X-ray map seen by XMM-Newton and in all synthesized \g-ray images. 
If the future reports of the H.E.S.S. collaboration reveal that the profile of VHE \g-ray 
brightness in SN~1006 closely match the XMM-Newton X-ray profile, then Fig.~\ref{ICpred:profiles} could 
suggests that the preferred model for SN~1006 could be MF2, i.e. the bright limbs in this SNR might be 
polar caps \citep{Rotetal04} in classical MHD description\footnote{Note, that the polar-caps model does not explain the azimuthal profiles of the radio brightness, if ISMF is uniform \citep{SN1006mf}.}. However, it seems that 
statistical errors affecting modern H.E.S.S. data may not allow us to distinguish between 
the three models of magnetic field. 
We expect however that direct comparison of the X-ray and \g-ray observations with more sofisticated 
simulations (e.g. involving nonuniform ISMF) allows one to find a correct three-dimensional geometry for 
SN~1006 and ISMF around it.

\section{Conclusions}

We propose a general method  to synthesize an  IC \g-ray image of  SNRs 
from  their  observed radio map and the 
results of spatially resolved X-ray spectral analysis, 
and we apply it to the supernova remnant SN~1006.  
The method is based on the fact that the surface brightness 
distribution of the synchrotron radio and X-ray emission of  SNRs  contains 
information about the distribution and properties of accelerated electrons which 
are responsible for the \g-ray emission as well. 
We have derived analytical expression to calculate the \g-ray image from 
radio or X-ray surface brightness map, Eq.~(\ref{ICpred:eq7}). 
It is used for each pixel of  the  input  image, 
knowing the  surface distribution of $\nu\rs{break}$ and assuming some 
configuration of magnetic field. 

X-ray and TeV \g-ray photons are radiated by electrons with almost the same 
energies. It seems therefore to be more natural to use the hard X-ray map as 
input rather than the radio one. However, we show that the contribution from the 
thermal component in the initial X-ray image may result in incorrect 
prediction for the IC brightness distribution. 
Thus, the hard X-ray map may be used  only if it is completely 
dominated everywhere by the nonthermal emission, which is not the case of
the XMM-Newton 2.0-4.5 keV image of SN~1006. 

On the other hand, the usage of the radio map as input may 
introduce some errors due to possible change of the spectral index in 
the  electron spectrum. The applicability of the method may be tested 
by  generating a  synchrotron X-ray image from the radio one with 
Eq.~(\ref{ICpred:eq4}). If synthesized distribution of X-ray surface brightness 
correlates well with the observed one (as in the case of SN~1006) 
then both the {\slshape srcut} model in 
XSPEC and our method for IC \g-ray image may be used for a given SNR. 

Our synthesized VHE IC \g-ray  
images of SN~1006 are quite similar to that reported in the publication 
of the H.E.S.S. collaboration \citep{HESSproc2009}.
This fact favours a leptonic scenario for the TeV \g-ray emission of this SNR.
However a hadronic origin cannot be ruled out in view of the measured ISM densities, 
consistent with a hadronic scenario \citep[e.g.][]{BVK2005}.
If this is the case, the observed TeV brightness map 
may reflect the distribution of protons with energies $>2.4\un{TeV}$ which interact with compressed 
ISM downstream of the shock. 

The present spatial resolution achieved by H.E.S.S., prevents us to ultimately disentangle
between the 
three considered configurations of MF in SN~1006: in all three cases, the two 
arcs on our \g-ray images are in the same location as in radio and X-rays. 
If MF strength varies no more than 
factor $\sim 3$ or so (as on Fig.~\ref{ICpredict:figMF}) around the shocks, 
IC \g-ray image of SN~1006 should have two limbs located 
in the same regions as in X-ray map, independently of the actual azimuthal 
configuration of MF. 
The reason, in accordance with Eq.~(\ref{ICpred:eq7}), is the 
contrasts in (radio or X-ray) brightness and $\nu\rs{break}$ which dominate 
any moderate azimuthal variation of MF. 
The use of the observed ratios of the radio surface brightness and 
the break frequency between NE and SE regions in Eq.~(\ref{ICpred:eq7}) shows that  
a variation of MF $B\rs{NE}/B\rs{SE}$ larger than a factor of 4 may reverse 
the location of bright IC limbs.

\section*{Acknowledgments}
O.P. acknowledge Osservatorio Astronomico di Palermo for hospitality. 
This work was partially supported by the program 'Kosmomikrofizyka' of National 
Academy of Sciences (Ukraine). 
D.I. acknowledges also support from the Program of Fundamental Research of the Physics 
and Astronomy Division of NASU. 
This work makes use of results produced by
the PI2S2 Project managed by the Consorzio COMETA, a project co-funded
by the Italian Ministry of University and Research (MIUR) within the
Piano Operativo Nazionale ``Ricerca Scientifica, Sviluppo Tecnologico,
Alta Formazione'' (PON 2000-2006). More information is available at
http://www.consorzio-cometa.it.  G.D. and G.C. acknowledge Argentina
grants from ANPCYT, CONICET, and UBA. 



\appendix
\section[]{Approximate description of surface brightness}

Here, the approach developed in \citet{SN1006mf,thetak} is adopted for general description of the surface brightness.
The surface brightness of a spherical SNR projection at distance $\varrho$ from the center and at azimuth $\varphi$ is 
\begin{equation}
 S(\varrho,\varphi)=2\int^{R}_{a(\varrho)}q(a,\Theta\rs{o}) {r r\rs{a} d a\over 
 \sqrt{r^2-\varrho^2}}.
 \label{icp:ap1-br}
\end{equation}
where $q$ is emissivity, $\Theta\rs{o}=\Theta\rs{o}(\varphi,r/\varrho,\phi\rs{o})$ is the shock obliquity, $\phi\rs{o}$ an aspect angle, $r$ and $a$ are Eulerian and Lagrangian coordinates, $r\rs{a}$ the derivative of $r(a)$ in respect to $a$. The emissivity in synchrotron or IC process is
\begin{equation}
 q=\int dE N(E)p(E,\nu).
\end{equation}
In the $\delta$-function approximation of the single-electron emissivity $p(E,\nu)$, we may write that 
\begin{equation}
 q\propto N(E\rs{m})B^\mathrm{x}
\end{equation}
where $E\rs{m}$ is an energy of electron which gives maximum contribution to radiation at a given frequency $\nu$, $\mathrm{x}=1/2$ for synchrotron and $\mathrm{x}=0$ for IC emission. 

Energy spectrum of electrons $N(E)$ evolves in a different way downstream of the shocks with different obliquity, i.e. $N(E\rs{m})=N(E\rs{m};a,\Theta\rs{o})$. In Sedov SNR, this evolution may approximately be expressed by the two independent terms \citep[for details see][]{SN1006mf,thetak}
\begin{equation}
 N(E\rs{m};a,\Theta\rs{o})\approx N\rs{a}(a)N\rs{\Theta}(\varrho,\Theta\rs{eff})
\end{equation}
where $\Theta\rs{eff}=\Theta\rs{o}(\varphi,1,\phi\rs{o})$. The similar relation holds for MF: 
$B(a,\Theta\rs{o})\approx B\rs{s}(\Theta\rs{o})B\rs{a}(a)$ where $B\rs{s}$ is the immediately post-shock value.
This allows Eq.~(\ref{icp:ap1-br}) to be written as 
\begin{equation}
\begin{array}{ll}
 S(\varrho,\varphi)&\propto \displaystyle
 N\rs{\Theta}(\varrho,\Theta\rs{eff})B\rs{s}(\Theta\rs{eff})^\mathrm{x}\\ \\
 &\displaystyle
 \times\int^{R}_{a(\varrho)}\! N\rs{a}(a)B\rs{a}(a)^\mathrm{x}
 {r r\rs{a} d a\over \sqrt{r^2-\varrho^2}}.
\end{array} 
 \label{icp:ap2}
\end{equation}
where integral depends on $\varrho$ only. 
In other notations, 
\begin{equation}
 S(\varrho,\varphi)\approx q\rs{eff}(\varrho,\varphi)\cdot {\cal I}(\varrho)
\end{equation}
where ${\cal I}$ is an integral in (\ref{icp:ap2}) devided by $N\rs{a}(\varrho)$. The accuracy of this approximate formula increases toward the edge of SNR projection where the bright limbs we are interested in are located. 

It is important that the factor ${\cal I}$ does not depend on $E\rs{m}$, but only on the coordinate in the projection.  
This means that ${\cal I}$ is almost the same in a given position of the radio, X- and $\gamma$-ray images and we may use Eqs.~(\ref{ICpred:eq4}), (\ref{ICpred:eq7}) and (\ref{ICpredeq8}) written for emissivities in order to relate surface brightnesses in each `pixel'. 
The factor ${\cal I}$ is also independent of $\varphi$ in this approximation. Therefore, 
the azimuthal variations of the surface brightness $S$ at a given $\varrho$ may be directly represented by the 
azimuthal variations of the effective emissivities. This provides justification for discussion in Sect.~\ref{icp:discussion}. However, the radial contrasts in brightness should account for the radial changes in ${\cal I}$ which is unknown until one considers detailed 3-D MHD model of SNR. 

\label{lastpage}

\end{document}